\documentclass[12pt,a4paper,final]{iopart}
\usepackage{iopams}
\usepackage{bm}

\expandafter\let\csname equation*\endcsname\relax
\expandafter\let\csname endequation*\endcsname\relax

\usepackage{amsmath}
\usepackage{amssymb}
\usepackage{amsthm}
\usepackage[utf8]{inputenc}

\usepackage[breaklinks=true,colorlinks=true,linkcolor=blue,urlcolor=blue,citecolor=blue]{hyperref}

\newcommand{\defeq}{\mathrel{\mathop:}=}

\begin{document}

\title{Conditional maximum entropy and Superstatistics}

\author[cor1]{Sergio Davis$^{1,2}$}
\address{$^1$Comisi\'on Chilena de Energía Nuclear, Casilla 188-D, Santiago, Chile}
\address{$^2$Departamento de F\'isica, Facultad de Ciencias Exactas, Universidad Andres Bello. Sazi\'e 2212, piso 7, 8370136, Santiago, Chile.}
\ead{sergio.davis@cchen.cl}

\begin{abstract}
Superstatistics describes nonequilibrium steady states as superpositions of canonical ensembles with a probability distribution of temperatures.
Rather than assume a certain distribution of temperature, recently [J. Phys. A: Math. Theor. \textbf{53}, 045004 (2020)] we have discussed general conditions
under which a system in contact with a finite environment can be described by superstatistics together with a physically interpretable, microscopic definition of
temperature. In this work, we present a new interpretation of this result in terms of the standard maximum entropy principle (MaxEnt) using conditional expectation
constraints, and provide an example model where this framework can be tested.
\end{abstract}

\section{Introduction}

The maximum entropy principle (MaxEnt)~\cite{Jaynes1957} provides a powerful explanation, deeply rooted in information theory, for the wide success of Boltzmann-Gibbs (BG)
statistical mechanics in equilibrium systems. Despite all this success, there are systems for which non-canonical steady states are the norm, a fact which has proved difficult to
reconcile with BG statistics. This has led to alternative frameworks such as nonextensive (Tsallis) statistics~\cite{Tsallis2009} and superstatistics~\cite{Beck2003, Beck2004}, among
others. While Tsallis statistics departs from the Boltzmann-Gibbs entropy by proposing a generalization of the entropic functional, superstatistics augments the maximum entropy principle
by providing a mechanism that produces non-canonical distributions without the need to abandon the foundations of Boltzmann-Gibbs statistics. This is achieved by postulating a
superposition of canonical ensembles, governed by a continuous or discrete distribution of \emph{inverse temperature} $\beta=1/k_B T$. More precisely, superstatistics produces ensembles
of the form
\begin{equation}
P(\bm x|S) = \rho(H(\bm x)),
\label{eq_ensemble}
\end{equation}
where $\bm{x}$ represents the system microstate, $H$ is the Hamiltonian function and $\rho(E)$ is the \emph{ensemble function} (also referred to as the generalized Boltzmann factor),
given by
\begin{equation}
\rho(E) = \int d\beta f(\beta)\exp(-\beta E).
\label{eq_rho}
\end{equation}

The traditional interpretation of superstatistics invokes models with a fluctuating parameter $\beta$ that follows some predefined stochastic dynamics~\cite{Reynolds2003, Beck2005,
Beck2007, Metzler2020}. Besides this, a Bayesian interpretation~\cite{Sattin2006,Sattin2018,Davis2018} of superstatistics is possible in which the superposition of temperatures can
be understood as a manifestation of the uncertainty in the value of the mean energy used as constraint in a maximum entropy setting. In this case, Eqs. \ref{eq_ensemble} and
\ref{eq_rho} are written, using the marginalization rule of probability~\cite{Jaynes2003,Sivia2006}, as
\begin{align}
P(\bm x|S) & = \int d\beta P(\beta|S)P(\bm x|\beta) \nonumber \\
           & = \int d\beta P(\beta|S) \left[\frac{\exp(-\beta H(\bm x))}{Z(\beta)}\right],
\label{eq_super_bayes}
\end{align}
under the equivalence $f(\beta) = P(\beta|S)/Z(\beta)$.

Recently\cite{Davis2020}, a mechanism leading to superstatistics has been explored where the original (target) system with microstates $\bm{x}$ is in contact with an environment with
microstates $\bm{y}$ and Hamiltonian $G(\bm y)$, in such a way that Eq. \ref{eq_super_bayes} is recovered by integration over the microstates of the environment, that is,
\begin{equation}
P(\bm x|S) = \int d\bm{y} P(\bm{x}, \bm{y}|S).
\label{eq_marg}
\end{equation}

\noindent
If we require that a \emph{microscopic inverse temperature} function $\mathcal{B}$ exists such that
\begin{equation}
\Big<g(\beta)\Big>_\mathcal{S} = \Big<g(\mathcal{B})\Big>_\mathcal{S}
\label{eq_megacond}
\end{equation}
for any function $g$ and that Eqs. \ref{eq_super_bayes} and \ref{eq_marg} are simultaneously true, then Ref.\cite{Davis2020} proves that this can be accomplished if and only if
the conditional distribution of the system given a fixed environment is canonical, with a temperature which is exclusively a property of the environment. That is,
\begin{equation}
P(\bm{x}|\bm{y}, S) = \left[\frac{\exp(-\beta H(\bm x))}{Z(\beta)}\right]\Big|_{\beta=\mathcal{B}(G(\bm y))}.
\label{eq_conditional}
\end{equation}

Here $\mathcal{B}=\mathcal{B}(G)$ is a function of the environment energy that is identified one-to-one with the superstatistical inverse temperature $\beta$, because Eq.
\ref{eq_megacond} with $g(\beta) = \delta(\beta-\beta_0)$ implies
\begin{equation}
P(\beta = \beta_0|S) = \Big<\delta(\mathcal{B}(G)-\beta_0)\Big>_S = P(\mathcal{B}=\beta_0|S).
\end{equation}

This treatment restores some aspects of the \emph{frequentist} intuition behind superstatistics by making $P(\beta|S)$ the sampling distribution of $\mathcal{B}$.
At the same time, it seems to lose some of the direct connection with the statistical inference frameworks, either Bayesian probability or MaxEnt. Furthermore, the condition
in Eq. \ref{eq_conditional} is strongly suggestive of an underlying maximum entropy formulation. This suggests the need in this formalism for some extra elements.

In this work we show that Eq. \ref{eq_conditional}, the neccesary and sufficient condition for superstatistics with a microscopic definition of \emph{environment temperature}, is
the natural consequence of a maximum entropy analysis using conditional expectations.

\newpage

This paper is organized as follows. First, in Section \ref{sect_maxent_cond}, we present the use of the standard maximum entropy principle in the case where the ratio of two
expectations is given as a constraint. In Section \ref{sect_maxent} we use this result to explore the effect of a conditional expectation of energy on the maximum entropy principle,
recovering Eq. \ref{eq_conditional}. Section \ref{sect_example} discusses an application of this formalism that can be solved exactly. Finally we close with some concluding remarks
in Section \ref{sect_conclud}.

\section{Maximum entropy with ratio constraints}
\label{sect_maxent_cond}

Before starting our main analysis, let us consider an important case in the maximum entropy formalism that is not commonly presented in the literature, but simple enough
to solve. Consider a constraint over the ratio of two expectations,
\begin{equation}
\frac{\big<f\big>}{\big<g\big>} = R,
\label{eq_ratio}
\end{equation}
where $f=f(\bm x)$ and $g=g(\bm x)$. The individual values of the expectations are unknown, only their ratio $R$ is assumed known. What is the most unbiased choice for $P(\bm x|R)$?

\noindent
According to MaxEnt, $P(\bm x|R)$ is equal to $p^*(\bm x)$, the function that maximizes the Boltzmann-Gibbs entropy
\begin{equation}
S[p] = -\int d\bm{x} p(\bm x) \ln p(\bm x)
\end{equation}
under the constraint on the ratio in Eq. \ref{eq_ratio} and normalization. This leads to the maximization of the augmented functional
\begin{equation}
\tilde{S} = -\int d\bm{x}p(\bm x)\ln p(\bm x) + \lambda\Big(\frac{\int d\bm{x}p(\bm x) f(\bm x)}{\int d\bm{x}p(\bm x)g(\bm x)}-R\Big) + \mu\Big(\int d\bm{x}p(\bm x) - 1\Big).
\end{equation}

\noindent
The equation defining the (unique) extremum of $\tilde{S}$, namely $p^*(\bm x)$, is then
\begin{equation}
\frac{\delta \tilde{S}}{\delta p(\bm x)}\Big|_{p=p^*} = 0 = -1 - \ln p^*(\bm x) + \lambda\frac{\delta}{\delta p(\bm x)}\left(\frac{\big<f\big>}{\big<g\big>}\right)\Big|_{p=p^*} + \mu.
\end{equation}

\noindent
Explicit solution of this problem requires the functional derivative of the ratio,
\begin{align}
\frac{\delta}{\delta p(\bm x)}\left(\frac{\big<f\big>}{\big<g\big>}\right) & = \frac{1}{\big<g\big>}\frac{\delta \big<f\big>}{\delta p(\bm x)} -
          \frac{\big<f\big>}{\big<g\big>^2}\frac{\delta \big<g\big>}{\delta p(\bm x)} \nonumber \\
      & = \frac{1}{\big<g\big>}\left(f(\bm x) - R\cdot g(\bm x)\right),
\end{align}
where we have used Eq. \ref{eq_ratio} to introduce $R$ in the second term of the right-hand side. We have then
\begin{equation}
\ln p^*(\bm x) = \frac{\lambda}{\big<g\big>}(f(\bm x)-R\cdot g(\bm x)) + \mu - 1.
\end{equation}

\noindent
At this point, as neither $\lambda$ nor $\big<g\big>$ are known, we can simply redefine $$\lambda \rightarrow -\frac{\lambda}{\big<g\big>},$$ and we have the maximum entropy solution
\begin{equation}
p^*(\bm x) = P(\bm x|R) = \frac{1}{Z(\lambda)}\exp\Big(-\lambda \big[f(\bm x) - R\cdot g(\bm x)\big]\Big),
\label{eq_ratio_solution}
\end{equation}
where the redefined Lagrange multiplier $\lambda$ is fixed by the constraint equation,
\begin{equation}
-\frac{\partial}{\partial \lambda}\ln Z(\lambda) = \big<f\big> - R\big<g\big> = 0.
\label{eq_ratio_lnZ}
\end{equation}

\noindent
Interestingly, the solution is the same as the naive use of the constraint in Eq. \ref{eq_ratio} in the form
\begin{equation}
\Big<f - R\cdot g\Big> = 0.
\end{equation}

\section{Conditional maximum entropy and superstatistics}
\label{sect_maxent}

Having the solution in Eqs. \ref{eq_ratio_solution} and \ref{eq_ratio_lnZ} to the ratio constraint in Eq. \ref{eq_ratio}, we are now equipped to present a
generalization of the textbook maximum entropy problem that leads to the canonical ensemble. Instead of the usual constraint on the mean energy,
\begin{equation}
\Big<H\Big>_S = \bar{E},
\label{eq_canon}
\end{equation}
where $\bar{E}$ is a known constant, we will consider that our system is placed in contact with an environment with Hamiltonian $G$ and we know the conditional mean value
of the energy, $\bar{E}(G_0)$ given that the environment is ``frozen'' at energy $G=G_0$, for $G_0 \in [G_\text{min}, G_\text{max}]$. Thus the constraint in Eq. \ref{eq_canon}
is generalized to
\begin{equation}
\Big<H\Big>_{S,G}\hspace{-2pt} = \bar{E}(G),
\label{eq_Econstr}
\end{equation}
for every value of $G \in [G_\text{min}, G_\text{max}]$. Just as in the canonical ensemble the value of $\bar{E}$ leads to a unique value of inverse temperature $\beta$,
the \emph{function} $\bar{E}(G)$ will lead to a distribution of inverse temperatures, as we will see shortly. Using the property
\begin{align}
\Big<H\cdot \delta(G-G_0)\Big>_S & = \Big<H\Big>_{S, G_0}\cdot P(G=G_0|S) \nonumber \\
                                        & = \bar{E}(G_0)\cdot \Big<\delta(G-G_0)\Big>_S
\end{align}
we can rewrite the constraint in Eq. \ref{eq_Econstr} as a ratio constraint similar to Eq. \ref{eq_ratio},
\begin{equation}
\frac{\Big<H\cdot \delta(G-G_0)\Big>_S}{\Big<\delta(G-G_0)\Big>_S} = \bar{E}(G_0), \qquad\;\;\;\;\forall\; G_0.
\end{equation}

\noindent
Hence the maximum entropy solution is given by
\begin{align}
P(\bm x, \bm y|S) & = \frac{1}{\eta}\exp\left(-\int_{G_\text{min}}^{G_\text{max}}\hspace{-2pt}dG_0 \lambda(G_0)\delta(G(\bm y)-G_0)\left[H(\bm x)-\bar{E}(G_0)\right]\right) \nonumber \\
                  & =  \frac{1}{\eta}\exp\left(-\lambda(G(\bm y))\left[H(\bm x)-\bar{E}(G(\bm y))\right]\right).
\label{eq_joint_P}
\end{align}

Now we show that this solution is compatible with Eq. \ref{eq_conditional}. We obtain first the marginal distribution of $\bm{y}$ by integrating over $\bm{x}$,
\begin{align}
P(\bm y|S) & = \int d\bm{x}P(\bm{x}, \bm{y}|S) \nonumber \\
           & = \frac{1}{\eta}\exp\left(\lambda(G(\bm y))\bar{E}(G(\bm y))\right)\times\left[\int d\bm{x} \exp\Big(-\lambda(G(\bm y))H(\bm x)\Big)\right] \nonumber \\
           & = \frac{1}{\eta}\exp\left(\lambda(G(\bm y))\bar{E}(G(\bm y))\right) \cdot Z(\lambda(G(\bm y))),
\label{eq_Py}
\end{align}
where $Z(\beta)$ is the partition function of the target system at inverse temperature $\beta$. By dividing $P(\bm x, \bm y|S)$ by $P(\bm y|S)$ we clearly obtain
\begin{equation}
P(\bm x|\bm y, S) = \frac{P(\bm x, \bm y|S)}{P(\bm y|S)} = \frac{\exp\left(-\lambda(G(\bm y))H(\bm x)\right)}{Z(\lambda(G(\bm y)))},
\end{equation}
but this is precisely Eq. \ref{eq_conditional} if we identify $\lambda(G)$ with $\mathcal{B}(G)$,
\begin{equation}
P(\bm x|\bm y, S) = \frac{\exp(-\beta H(\bm x))}{Z(\beta)}\Big|_{\beta = \mathcal{B}(G(\bm y))}.
\label{eq_Px_cond}
\end{equation}

\noindent
We determine the Lagrange multiplier $\mathcal{B}$ by imposing the constraint in Eq. \ref{eq_Econstr},
\begin{align}
\bar{E}(G) = \big<H\big>_{S,G} & = \int d\bm{x} P(\bm x|G, S)H(\bm x) \nonumber \\
            & = \int d\bm{x} H(\bm x) \int d\bm{y} P(\bm x|\bm y, S)P(\bm y|G, S) \nonumber \\
            & = \int d\bm{x} H(\bm x)\hspace{-5pt} \int d\bm{y} \left[\frac{\exp(-\beta H(\bm x))}{Z(\beta)}\right]\Big|_{\beta=\mathcal{B}(G(\bm y))}\hspace{-5pt} \times\left[\frac{\delta(G(\bm y)-G)}{\Omega_G(G)}\right],
\end{align}
where in the last line we have used Eq. \ref{eq_Px_cond} and the fact that $P(\bm y|S)$ only depends on $\bm y$ through $G(\bm y)$. Evaluating the integrals, we have
\begin{equation}
\bar{E}(G)  = \int d\bm{x}\; \frac{\exp(-\mathcal{B}(G) H(\bm x))H(\bm x)}{Z(\mathcal{B}(G))}
            = \left[-\frac{\partial}{\partial \beta}\ln Z(\beta)\right]\Big|_{\beta=\mathcal{B}(G)} = \varepsilon(\mathcal{B}(G)),
\label{eq_EG}
\end{equation}
where $\varepsilon(\beta)$ is the canonical caloric curve, given by
\begin{equation}
\varepsilon(\beta) \defeq \Big<H\Big>_\beta = -\frac{\partial}{\partial \beta}\ln Z(\beta).
\end{equation}

\noindent
Therefore, if $\varepsilon(\beta)$ is invertible, we have
\begin{equation}
\mathcal{B}(G) = \varepsilon^{-1}(\bar{E}(G)).
\end{equation}

The physical interpretation of this correspondence is that the microscopic inverse temperature $\mathcal{B}(G)$ is the one that, in the canonical caloric curve,
yields the correct mean energy $\big<H\big>_{\beta=\mathcal{B}} = \big<H\big>_{S,G}$ for every admissible value of $G$. Thus, in the target system, $\beta$ follows the
fluctuations of $G$ mapped through the canonical caloric curve.

The marginal distribution $P(\bm x|S)$ resulting from integration over $\bm{y}$ of Eq. \ref{eq_joint_P} will be described by superstatistics with $P(\beta|S) = P(\mathcal{B}=\beta|S)$,
which we can write in terms of our original quantity $\bar{E}(G)$. This gives
\begin{align}
P(\beta|S) & = \Big<\delta(\mathcal{B}(G)-\beta)\Big>_S \nonumber \\
           & = \int d\bm{y} \left[ \frac{1}{\eta}Z(\mathcal{B}(G(\bm y)))\exp\left(\mathcal{B}(G(\bm y))\bar{E}(G(\bm y))\right)\delta(\mathcal{B}(G(\bm y))-\beta)\right] \nonumber \\
           & = \frac{1}{\eta}Z(\beta)\int d\bm{y}\exp(\beta \bar{E}(G(\bm y)))\delta(\mathcal{B}(G(\bm y))-\beta) \nonumber \\
           & = \frac{1}{\eta}Z(\beta)\int dG \Omega_G(G)\exp(\beta \bar{E}(G))\delta(\mathcal{B}(G)-\beta).
\end{align}

\noindent
By using the following property of the Dirac delta function,
\begin{equation}
\delta(\mathcal{B}(G)-\beta) = \sum_{G^*} \frac{\delta(G^*-G)}{|\mathcal{B}'(G)|},
\end{equation}
where $G^*$ are the solutions of $\mathcal{B}(G^*) = \beta$, we derive the explicit formula
\begin{equation}
P(\beta|S) = \frac{1}{\eta}Z(\beta)\sum_{G^*} \frac{\Omega_G(G^*)}{|\mathcal{B}'(G^*)|}\exp(\beta\bar{E}(G^*)).
\label{eq_Pbeta_many}
\end{equation}
In order to determine $G^*$, we use Eq. \ref{eq_EG}, which gives us the connection between $\bar{E}$ and $\mathcal{B}$, $$\bar{E}(G) = \varepsilon(\mathcal{B}(G)).$$

\noindent
In the case of $G=G^*$, this becomes $\bar{E}(G^*) = \varepsilon(\mathcal{B}(G^*)) = \varepsilon(\beta)$ and so, if the relation $\bar{E}(G)$ is invertible, we have
\begin{equation}
G^*(\beta) = \bar{E}^{-1}(\varepsilon(\beta)),
\label{eq_Gstar}
\end{equation}
and we can obtain our main result,
\begin{equation}
P(\beta|S) = \frac{1}{\eta}Z(\beta)\Omega_G(G^*(\beta))\Big|\frac{\varepsilon'(\beta)}{\bar{E}'(G^*(\beta))}\Big|\exp(\beta\varepsilon(\beta)),
\label{eq_Pbeta}
\end{equation}
with $G^*$ given by Eq. \ref{eq_Gstar}. In terms of the usual notation of superstatistics,
\begin{equation}
f(\beta) = \frac{1}{\eta}\Omega_G(G^*(\beta))\Big|\frac{\varepsilon'(\beta)}{\bar{E}'(G^*(\beta))}\Big|\exp(\beta\varepsilon(\beta)),
\label{eq_fbeta}
\end{equation}
and the corresponding ensemble function $\rho(E)$ for the target system is
\begin{equation}
\rho(E) = \frac{1}{\eta}\int d\beta \Omega_G(G^*(\beta))\Big|\frac{\varepsilon'(\beta)}{\bar{E}'(G^*(\beta))}\Big|\exp(-\beta[E-\varepsilon(\beta)]).
\label{eq_rhoE}
\end{equation}

\section{Example}
\label{sect_example}

Let us demonstrate this formalism with a concrete example. Consider the target and environment in such a state that
\begin{equation}
\Big<H + G\Big>_{S,G_0} = \big<H\big>_{S, G_0} + G_0 = E_0,
\label{eq_example_fixed}
\end{equation}
with $E_0$ the fixed total energy of the system. Furthermore, let us assume the densities of states
\begin{align}
\Omega(E) &= c_0 E^\alpha, \nonumber \\
\Omega_G(G) & = c_1 \exp(\beta_0 G)
\label{eq_example_dens}
\end{align}
for the target system and the environment, respectively. From the density of states of the target we can obtain its
partition function,
\begin{equation}
Z(\beta) = \int dE\;\Omega(E)\exp(-\beta E) = c_0\Gamma(\alpha+1)\beta^{-(\alpha+1)},
\end{equation}
as well as its canonical caloric curve,
\begin{equation}
\varepsilon(\beta) = -\frac{\partial}{\partial \beta}\ln Z(\beta) = \frac{\alpha+1}{\beta}.
\label{eq_ex_caloric}
\end{equation}

\noindent
From Eq. \ref{eq_example_fixed} we can obtain $\bar{E}(G)$ as
\begin{equation}
\bar{E}(G) = \big<H\big>_{S, G} = E_0 - G,
\label{eq_ex_barE}
\end{equation}
and replacing Eqs. \ref{eq_ex_barE} and \ref{eq_ex_caloric} into Eq. \ref{eq_EG}, we obtain
\begin{equation}
\bar{E}(G) = E_0 - G = \varepsilon(\mathcal{B}(G)) = \frac{\alpha+1}{\mathcal{B}(G)},
\end{equation}
from which we read the microscopic inverse temperature
\begin{equation}
\mathcal{B}(G) = \frac{\alpha+1}{E_0 - G}
\label{eq_example_beta}
\end{equation}
that the environment imposes over the system. Because the energy $G$ of the environment is such that $G \in [0, E_0]$, we have $\mathcal{B} \in [\beta_\text{min}, \infty)$,
where we have defined
\begin{equation}
\beta_\text{min} \defeq \frac{\alpha + 1}{E_0}.
\end{equation}

\noindent
This means the microscopic temperature $T(G) = 1/(k_B \mathcal{B}(G))$ is bounded from above, $$0 \leq T(G) \leq \frac{E_0}{k_B (\alpha+1)}.$$ Moreover, because the function
$\bar{E}(G)$ is invertible, we have
\begin{equation}
G^*(\beta) = E_0 - \frac{\alpha+1}{\beta} = E_0\Big(1-\frac{\beta_\text{min}}{\beta}\Big).
\end{equation}

In order to make the statistical treatment more smooth, let us consider the limit where $E_0$ is large enough that we can take $\beta \in [0, \infty)$ and $E \in [0, \infty)$.
Now we have everything to compute the statistical properties. First, we compute the joint energy-inverse temperature distribution,
\begin{align}
P(E, \beta|S) & = P(E|\beta)\times P(\beta|S) \nonumber \\
              & = \exp(-\beta E)\Omega(E)f(\beta) \nonumber \\
              & = \frac{[\beta_0(\alpha+1)]^{\alpha+2}}{\Gamma(\alpha+1)\Gamma(\alpha+2)}\exp(-\beta_0(\alpha+1)/\beta)\beta^{-2}\exp(-\beta E)E^\alpha.
\label{eq_PEbeta}
\end{align}

\noindent
The marginal distribution of $\beta$, equivalent to Eq. \ref{eq_Pbeta}, yields
\begin{equation}
P(\beta|S) = \frac{[\beta_0(\alpha+1)]^{\alpha+2}}{\Gamma(\alpha+2)}\beta^{-(\alpha+3)}\exp(-\beta_0(\alpha+1)/\beta),
\end{equation}
that is, an inverse gamma distribution, which means the temperature $T = 1/(k_B \beta)$ that the target system ``sees'' follows a gamma distribution. From this distribution
we can compute the mean inverse temperature
\begin{equation}
\big<\beta\big>_S = \beta_0
\end{equation}
and its normalized variance,
\begin{equation}
\frac{\Big<(\delta \beta)^2\Big>_S}{\big<\beta\big>_S^2} = \frac{1}{\alpha},
\label{eq_beta_var}
\end{equation}
and we see that $\alpha \rightarrow \infty$ keeping $\beta_0$ fixed leads to $\big<(\delta \beta)^2\big>_S \rightarrow 0$, recovering the canonical ensemble with $\beta=\beta_0$.
The ensemble function $\rho(E)$ of the target system is
\begin{equation}
\rho(E) = \frac{2}{c_0}\frac{[\beta_0(\alpha+1)]^{\alpha+1}}{\Gamma(\alpha+1)\Gamma(\alpha+2)}\sqrt{\beta_0(\alpha+1)E}K_1(2\sqrt{\beta_0(\alpha+1)E}),
\end{equation}
where $K_1(z)$ is the modified Bessel function of the second kind, and the corresponding energy distribution of the target is
\begin{equation}
P(E|S) = \rho(E)\Omega(E) = \frac{2\beta_0(\alpha+1)}{\Gamma(\alpha+1)\Gamma(\alpha+2)}[\beta_0(\alpha+1)E]^{\alpha+\frac{1}{2}}K_1(2\sqrt{\beta_0(\alpha+1)E}).
\end{equation}

\noindent
The mean target energy is
\begin{equation}
\big<E\big>_S = \frac{\alpha+2}{\beta_0},
\end{equation}
while its normalized variance is
\begin{equation}
\frac{\Big<(\delta E)^2\Big>_S}{\big<E\big>_S^2} = \frac{2}{\alpha+1}.
\label{eq_E_var}
\end{equation}

\newpage

In this superstatistical ensemble described by Eq. \ref{eq_PEbeta}, the fluctuations of (inverse) temperature and energy are well defined, and we can explore what their
joint behavior is. As we can see from Eqs. \ref{eq_beta_var} and \ref{eq_E_var}, when $\alpha$ increases with fixed $\beta_0$, the variance of $\beta$ tends to zero
as $1/\alpha$, while the variance of $E$ increases linearly with $\alpha$, so their product remains constant for large enough $\alpha$.

In fact, it is possible to construct a \emph{thermodynamic uncertainty relation} for the \emph{unnormalized} variances of $\beta$ and $E$ in the target system, as have been proposed
and studied in previous works~\cite{Uffink1999,Velazquez2009d}, namely
\begin{equation}
\Big<(\delta \beta)^2\Big>_S\cdot \Big<(\delta E)^2\Big>_S = 2\left[\frac{(\alpha+2)^2}{\alpha(\alpha+1)}\right] \geq 2.
\end{equation}

Here the variance of the intensive parameter $\beta$ can in fact be interpreted as fluctuations of a physical quantity, which decrease when the fluctuations of energy
increase. On the other hand, we can compare this uncertainty relation with the \emph{correlation} between $\beta$ and $E$ that we can obtain from the joint distribution,
\begin{equation}
\Big<\delta \beta\cdot \delta E\Big>_S = \big<\beta E\big>_S - \big<\beta\big>_S\big<E\big>_S = (\alpha+1) - (\alpha+2) = -1,
\end{equation}
in direct agreement with Schwarz inequality, $\big<(\delta \beta)^2\big>_S\cdot\big<(\delta E)^2\big>_S \geq \big<\delta \beta\cdot \delta E\big>^2$.

\section{Concluding remarks}
\label{sect_conclud}

We have complemented the framework initiated in Ref.\cite{Davis2020}, by connecting it with the maximum entropy principle under a conditional energy expectation
constraint, $\big<H\big>_{S,G} = \bar{E}(G)$. In this formalism, knowledge of $\bar{E}(G)$, the canonical caloric curve $\varepsilon(\beta)$ and the density of
states $\Omega_G$ of the environment completely determines the \emph{most unbiased} form of superstatistics, with $f(\beta)$ or $\rho(E)$ given by Eqs. \ref{eq_Gstar},
\ref{eq_fbeta} and \ref{eq_rhoE}. The frequentist interpretation of $P(\beta|S)$ as a sampling distribution of an observable of the environment remains valid.
We have explored a simple model of target and environment for which one can obtain fluctuations of the intensive parameter $\beta$ in a manner consistent with a
frequentist interpretation of superstatistics, and also consistent with the Bayesian/MaxEnt framework, only requiring the standard elements of traditional MaxEnt and
the laws of probability.

\section*{Acknowledgements}

This work is supported by the Anillo ACT-172101 grant.

\newpage

\section*{References}

%\bibliography{maxent}

\begin{thebibliography}{10}

\bibitem{Jaynes1957}
E.~T. Jaynes.
\newblock Information theory and statistical mechanics.
\newblock {\em Phys. Rev.}, 106:620--630, 1957.

\bibitem{Tsallis2009}
C.~Tsallis.
\newblock {\em Introduction to nonextensive statistical mechanics: approaching
  a complex world}.
\newblock Springer Science \& Business Media, 2009.

\bibitem{Beck2003}
C.~Beck and E.G.D. Cohen.
\newblock Superstatistics.
\newblock {\em Phys. A}, 322:267--275, 2003.

\bibitem{Beck2004}
C.~Beck.
\newblock Superstatistics: theory and applications.
\newblock {\em Cont. Mech. Thermodyn.}, 16:293--304, 2004.

\bibitem{Reynolds2003}
A.~M. Reynolds.
\newblock Superstatistical mechanics of tracer-particle motions in turbulence.
\newblock {\em Phys. Rev. Lett.}, 91:084503, 2003.

\bibitem{Beck2005}
C.~Beck, E.~G.~D. Cohen, and H.~L. Swinney.
\newblock From time series to superstatistics.
\newblock {\em Phys. Rev. E}, 72:056133, 2005.

\bibitem{Beck2007}
C.~Beck.
\newblock Statistics of three-dimensional {L}agrangian turbulence.
\newblock {\em Phys. Rev. Lett.}, 98:064502, 2007.

\bibitem{Metzler2020}
R.~Metzler.
\newblock Superstatistics and non-gaussian diffusion.
\newblock {\em Eur. Phys. J.: Special Topics}, 229:711--728, 2020.

\bibitem{Sattin2006}
F.~Sattin.
\newblock Bayesian approach to superstatistics.
\newblock {\em Eur. Phys. J. B}, 49:219--224, 2006.

\bibitem{Sattin2018}
F.~Sattin.
\newblock Superstatistics and temperature fluctuations.
\newblock {\em Phys. Lett. A}, 382:2551--2554, 2018.

\bibitem{Davis2018}
S.~Davis and G.~Gutiérrez.
\newblock Temperature is not an observable in superstatistics.
\newblock {\em Phys. A}, 505:864--870, 2018.

\bibitem{Jaynes2003}
E.~T. Jaynes.
\newblock {\em Probability Theory: The Logic of Science}.
\newblock Cambridge University Press, 2003.

\bibitem{Sivia2006}
D.~S. Sivia and J.~Skilling.
\newblock {\em Data Analysis: A Bayesian Tutorial}.
\newblock Oxford University Press, 2006.

\bibitem{Davis2020}
S.~Davis.
\newblock On the possible distributions of temperature in nonequilibrium steady
  states.
\newblock {\em J. Phys. A: Math. Theor.}, 53:045004, 2020.

\bibitem{Uffink1999}
J.~Uffink and J.~Van Lith.
\newblock Thermodynamic uncertainty relations.
\newblock {\em Found. Phys.}, 29:655--692, 1999.

\bibitem{Velazquez2009d}
L.~Velazquez and S.~Curilef.
\newblock A thermodynamic fluctuation relation for temperature and energy.
\newblock {\em J. Phys. A: Math. Theor.}, 42:95006, 2009.

\end{thebibliography}
%\bibliographystyle{unsrt}

\end{document}